# Tunable Directional Filter for Mid-Infrared Optical Transmission Switching

ANDREW BUTLER, JACK SCHULZ, AND CHRISTOS ARGYROPOULOS[*]

*Department of Electrical and Computer Engineering, University of Nebraska-Lincoln, Lincoln, Nebraska 68588, USA*
*\*christos.argyropoulos@unl.edu*

**Abstract:** Controlling the spectral and angular response of infrared (IR) radiation is a challenging task of paramount importance to various emerging photonic applications. Here, we overcome these problems by proposing and analyzing a new design of a tunable narrowband directional optical transmission filter. The presented thermally controlled multilayer filter leverages the temperature dependent phase change properties of vanadium dioxide ($VO_2$) to enable efficient and reversible fast optical switching by using a pump-probe laser excitation setup. More specifically, transmission is blocked for high intensity probe lasers due to the $VO_2$ metallic properties induced at elevated temperatures while at low probe laser intensities high transmission through the filter occurs only for a narrowband IR range confined to near normal incident angles. The proposed multilayer composite dielectric filter is expected to have applications in optical communications, where it can act as dual functional infrared filter and optical switch.

## 1. Introduction

Recently, the emerging area of tunable photonics has attracted increased attention as an efficient way to design new optical devices that can dynamically adapt their properties to a variety of diverse applications. A particular phase change material commonly used in tunable photonic applications is vanadium dioxide ($VO_2$). It undergoes a reversible phase transition from insulator to metal as its temperature is increased by switching phases at a critical temperature (~340K) [1], which can be controlled and tuned via doping with tungsten [2]. Thus, $VO_2$ enables efficient thermal control over the optical properties of photonic structures that is incorporated into. Many works were focused on using $VO_2$ to thermally switch between states of high and low reflection, transmission, or absorption. As an example, tunable reflection in the visible spectrum was demonstrated by using $VO_2$ as a defect layer in a Bragg filter design [3]. Enhanced tunability of absorption/emissivity in the mid-infrared (IR) band was demonstrated using a multilayer design composed of alternating layers of $VO_2$ and metals [4]. Another design utilized the refractive index change of $VO_2$ to create a tunable optical response in a thin film structure [5]. A $VO_2$ filled double cavity was designed for operation in the near IR that could act as band-pass filter or optical switch [6]. More recently, switchable transmission in the mid-IR range was demonstrated using a thin $VO_2$ layer in a multilayer filter design [7]. However, most of the previously works demonstrate switching and reconfigurability only by varying the bulk temperature of the obtained device and not by dynamically changing the incident laser power in a pump-probe laser experimental configuration setup. The latter approach will be much faster in terms of switching operation since laser-induced heating can occur in extremely fast time scales [8].

Another important aspect of light propagation that the works above do not consider is filtering the incident angle of light that can be narrowed to a small range leading to directional transmission response. For full control of light, it is desirable for optical filters to not only control the spectral response but also the spatial directionality of incident light. In the past, angular filtering of light was demonstrated using a multilayer structure surrounded by a fluid serving as broadband impedance matching to the surrounding material [9]. This multilayer filter design featured efficient and broadband transmission restricted to only angles near 60°, however the design only worked for transverse-magnetic (TM) polarized light which limits its thermal emission applications. A narrowband polarization independent filter that confines the transmission of light to only near normal incidence has also been demonstrated in recent years [10]. While both of these designs perform well in terms of simultaneous spectral and angular filtering, they have no mechanisms to control or modulate their behavior which will be very beneficial for tunable photonic filter applications.

As mentioned before, another interesting application for tunable filters is optical switching based on pump-probe laser experiments. The binary switching of $VO_2$ between two phases makes it a good candidate for use in various switching scenarios. Optical switching using $VO_2$ thin films [11,12] and fast switching using $VO_2$ nanocrystals has been demonstrated for operation at near-IR frequencies [13]. Other designs have attempted to improve the modulation depth in the near-IR range, such as structures consisting of gold nanowires over a $VO_2$ thin spacer layer that leverages the plasmonic modes of metallic $VO_2$ to enhance tunability [14]. In the mid-IR range, another design used a two-dimensional (2D) array of aluminum nanoantennas over a thin $VO_2$ layer that alternated between high reflection and absorption depending on the $VO_2$ phase [15]. However, both these works require complex nanostructuring which makes their fabrication challenging and impractical.

Beyond $VO_2$, several alternative methods exist to create tunable filters and optical switches. Graphene is one common approach, where the Fermi energy of graphene can be tuned via electrostatic gating. This allows for control over the plasmonic resonances of graphene enabling tunable optical filters consisting of graphene layers combined with plasmonic gratings [16,17]. However, the use of graphene can result in additional fabrication complexity. Other techniques to create optical switches include controlling the position of liquid droplets [18], or the use of liquid crystals [19], Dirac semimetals [20], and topological semimetals based on antimonene [21]. A notable advantage of $VO_2$ over the other approaches is that the extreme difference between its metallic and insulating phase enable very high modulation depths [15] that can occur in rapid time scales in case the phase transition is induced by probe lasers.

In this work, we present the design of a new optically switchable narrowband directional transmission filter based on a simple multilayer dielectric structure. We utilize the thermally dependent optical properties of $VO_2$ to introduce control in the directional filtering response. When the $VO_2$ layers are in the insulating phase, the structure exhibits very narrowband transmission of light at incident angles close to normal. In the metallic phase, the $VO_2$ layers prevent transmission through the structure and turn the device to a mirror or absorber of mid-IR radiation when high probe laser intensities are utilized. The optical response of the design is accurately analyzed using the transfer matrix method and the dynamic temperature dependence is characterized using the $VO_2$ Bruggeman effective medium theory. Transient modeling under pump-probe laser illumination is carried out to demonstrate the applicability of the presented filter to fast optical switching at the mid-IR range. The presented directional and rapid switching response is unique among other relevant optical filters and provides an additional degree of control compared to other designs. The proposed optical filter is envisioned to have a plethora of photonic applications, such as infrared filtering and optical switching.

## 2. Narrowband directional filter analysis

The schematic of the proposed multilayer directional transmission filter is shown in Fig. 1a. The design consists of five unit cells of alternating layers of $VO_2$ and calcium fluoride ($CaF_2$). The thickness of each $VO_2$ layer is 450nm while each $CaF_2$ layer is 2.55μm thick. Both $VO_2$ and $CaF_2$ materials are modelled using their frequency dispersive dielectric constants obtained from relevant experimental data [22–24]. In the insulating state, the $VO_2$ acts as a high refractive index dielectric. When combined with the low index $CaF_2$ material, a special Bragg mirror effect occurs that works for both polarizations, creating a band of high reflection at wavelengths above 7.7μm. At the edge of this reflection band, a transmission resonance occurs, and a highly directional response is obtained. In a typical Bragg reflector design, the layer thicknesses are chosen to be ¼ the center wavelength of the stop band. Using this as a starting point, the thicknesses of the layers were optimized until the highly narrowband directional transmission results were obtained. The multilayer structure can be feasibly fabricated using thin film deposition techniques. Growth of $VO_2$ thin films on the order of 100nm thick have been demonstrated using RF magnetron sputtering [25] and thicker high quality $VO_2$ films have recently been demonstrated using the vapor transport method [26]. Likewise, $CaF_2$ films of micron thickness have been deposited using resistive evaporation [27]. Alternating these deposition techniques can realize the filter's multilayer structure.

The computed transmittance versus wavelength is shown in Fig. 1b at the primary resonance of the structure. The full width at half maximum (FWHM) for this resonance is found to be ~240nm which further demonstrates that the presented response is very narrowband. The transmittance spectra as a function of incidence angles are shown in Fig. 1c for TM polarization and Fig. 1d for transverse-electric (TE) polarization when $VO_2$ is in the insulating phase. A broader wavelength range is shown that includes multiple higher order resonances. These results prove that the presented filter can work for both polarizations, i.e., incoherent light illumination, consisting ideal property for thermal imaging and sensing. In the metallic $VO_2$ phase, the transmission is zero for all wavelengths and incident angles and the filter exhibits mirror-like response for near-normal incidence, while higher angles of incidence show broadband absorption. More information on the reflectance and absorptance in the metallic state is available in the Supplementary Information [28]. Figure 1e demonstrates the directional response of the filter when $VO_2$ is in the insulating phase, where it is confirmed that the optical transmission is limited to only a small range of angles near normal incidence. Interestingly, the narrowband directional response is maintained for a reduced number of material layers without a drastic drop in performance, as discussed in more details at the Supplementary Information [28].

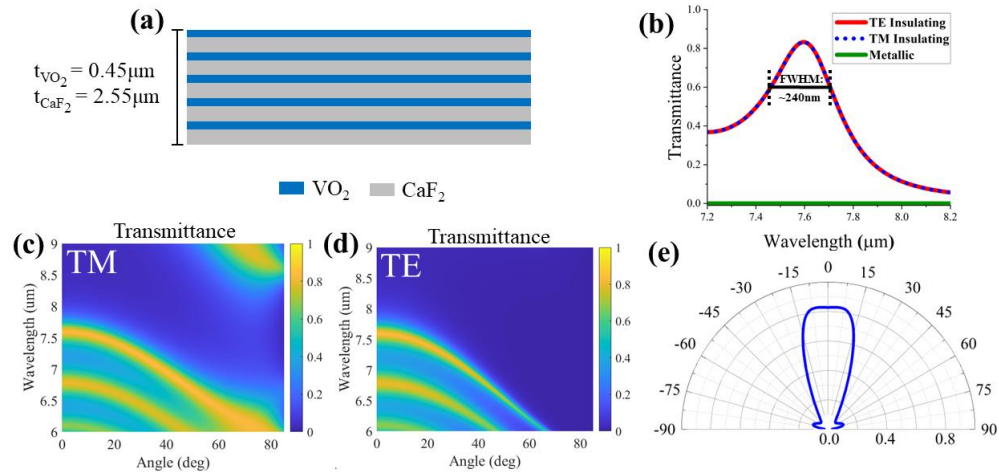

Fig 1. (a) Schematic of the multilayer directional transmission filter. (b) Transmittance as a function of wavelength for normal incidence illumination. Only the primary resonance is shown. There is substantial transmission change between insulating and metallic $VO_2$ phases. (c)-(d) Transmittance as a function of wavelength and incident angle for (c) TM and (d) TE polarization. A broader wavelength range containing higher order resonances is shown in these figures. (e) Directional transmittance pattern as a function of angle for fixed 7.6μm operation wavelength.

The transmittance spectra shown in Figs. 1b-e were obtained analytically using the transfer matrix method. Each layer was modelled as a transmission line and the ABCD parameters for each layer were obtained using [29]:

$$\begin{bmatrix} A & B \\ C & D \end{bmatrix} = \begin{bmatrix} \cos(kt) & jZ\sin(kt) \\ j\frac{1}{Z}\sin(kt) & \cos(kt) \end{bmatrix}, \qquad (1)$$

where t is the thickness of the layer, $k = k_0\sqrt{\varepsilon - \sin^2(\theta)}$ is the wavevector in the material, and $Z = \eta_0 / \sqrt{\varepsilon - \sin^2(\theta)}$ is the impedance for TE polarization, while the impedance becomes $Z = \eta_0\sqrt{\varepsilon - \sin^2(\theta)}$ for TM polarization. The impedance of free space is $\eta_0$ (377Ω), $k_0 = 2\pi/\lambda$, ε is the complex frequency dependent dielectric constant of each material, and θ is the incident angle. The total ABCD matrix for the whole multilayer structure can be obtained by multiplying the ABCD matrices of each layer:

$$\begin{bmatrix} A & B \\ C & D \end{bmatrix}_{tot} = \begin{bmatrix} A & B \\ C & D \end{bmatrix}_1 \begin{bmatrix} A & B \\ C & D \end{bmatrix}_2 \cdots \begin{bmatrix} A & B \\ C & D \end{bmatrix}_{10}. \qquad (2)$$

The transmission coefficient ($S_{21}$) through the structure can be found by converting the ABCD parameters of Eq. (2) into the S-parameters [29]:

$$S_{21} = \frac{2}{A + \frac{B}{Z_0} + CZ_0 + D}, \qquad (3)$$

where the free space impedance for oblique incident waves is $Z_0 = \eta_0 / \cos(\theta)$ for TE polarization and $Z_0 = \eta_0 \cos(\theta)$ for TM polarization. The transmittance through the structure is then found as $|S_{21}|^2$.

From the results in Figs. 1b-e, it can be derived that the transmission through the filter is very narrowband at ~7.6μm and is restricted to within ~15° from normal incidence. This small transmission band leads to narrow directional spatial filtering obtained at the edge of the multilayer structure's stopband. The highest transmission occurs closest to the stopband of the filter, though the fringes at lower IR wavelengths can also be used for similar performance filtering. The operating wavelength and fringe wavelengths can be adapted to different frequency ranges by simple redesigning the layer thicknesses. Interestingly, the presented filter design is polarization independent, thus the filter can be used with randomly polarized incoherent light sources, e.g., mid-IR thermal emission.

The directionality was further verified using finite element method 2D full wave simulations. An out of plane line current was used to model a point source located above the filter at a wavelength of 7.6μm. Again, the $VO_2$ and $CaF_2$ materials were modeled using their frequency dependent dielectric constants [22–24]. The induced field distributions through the filter for the $VO_2$ insulating and metallic phases are shown in Figs. 2a and 2b, respectively. Directional radiation is only present for the $VO_2$ insulating phase (Figs. 2a) while the transmission is terminated when the $VO_2$ is metallic (Figs. 2b). These results further prove the unique directionality and tunability of the filter design.

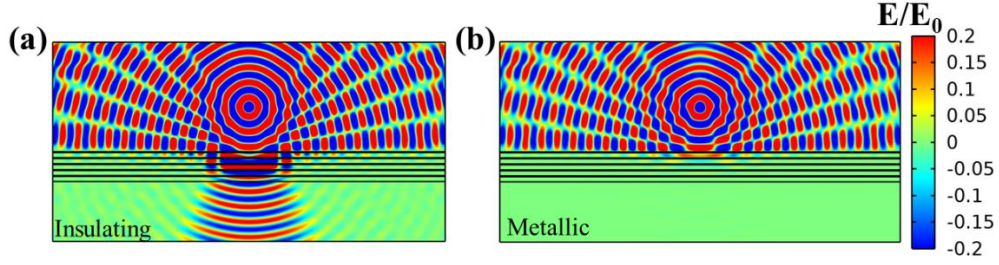

Fig 2: Field distributions through the presented directional filter using a point source when VO$_2$ is in the (a) insulating and (b) metallic phase, respectively.

The transmission of the presented filter does not change abruptly during the VO$_2$ phase transition process, but it has a dynamic temperature dependent response that follows the change of the VO$_2$ dielectric constant. More specifically, the VO$_2$ phase transition occurs at the critical temperature $(T_C \simeq 340K)$. Interestingly, the VO$_2$ phase is in transition mode at temperatures very close to critical, and its complex dielectric constant can be accurately computed by using the Bruggeman effective medium theory [30]:

$$\varepsilon_{VO_2} = \frac{1}{4}\left[\varepsilon_i(2-3V) + \varepsilon_m(3V-1) + \sqrt{\left[\varepsilon_i(2-3V) + \varepsilon_m(3V-1)\right]^2 + 8\varepsilon_i\varepsilon_m}\right], \qquad (4)$$

where $\varepsilon_m$ is the VO$_2$ dielectric constant in the metallic phase, $\varepsilon_i$ is the dielectric constant in the insulating phase [24], and V is the metallic volume fraction that can be calculated by using the formula:

$$V = 1 - \frac{1}{1+\exp(\frac{T-T_C}{\Delta T})}, \qquad (5)$$

where T is the ambient temperature, T$_C$ is the VO$_2$ critical temperature, and $\Delta T=2K$ [29] is the transition width. The formulas in Eqs. (4)-(5) can be used in the calculations of Eqs. (1)-(3) to accurately compute the dynamic temperature response of the filter. The transmission dependence on the temperature for normal incidence illumination at the optimal wavelength of 7.6μm is demonstrated in Fig. 3a. In addition, the angular response of the transmittance is shown in Fig. 3b as a function of temperature again at the fixed wavelength of 7.6μm.

The switching behavior of the optical filter as the temperature is increased is not abrupt but it has a dynamic transition range which is clearly illustrated in Fig. 3a. At lower temperatures than T$_C$, when VO$_2$ is in the insulating phase, transmission through the structure is greater than 80%, similar to Fig. 1b. At higher temperatures compared to T$_C$, VO$_2$ transitions to the metallic phase and transmission drops to zero. However, the directionality of the filter is maintained during the insulating and transition phases, as depicted in Fig. 3b. The relative abrupt but continuous transition between phases can enable fast optical transmission switching and modulation when the temperature rapidly changes. This type of response is ideal for efficient infrared optical communications, where the selective angular response of the filter can be used to eliminate noise and erroneous stray signals coming from oblique incident angles. Furthermore, the VO$_2$ critical temperature can be further reduced via doping with tungsten, thus the presented optical filter can be made tunable and even operate at room temperature.

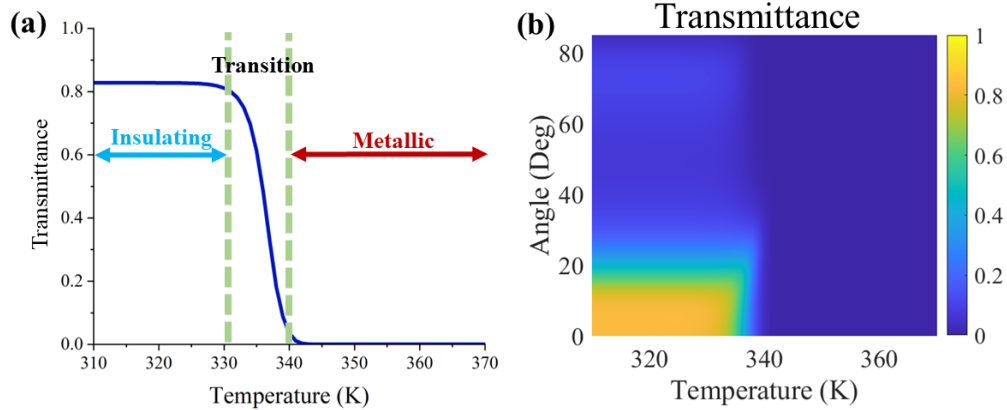
Fig 3. (a) Dynamic transmission response versus temperature at normal incidence illumination. The temperature ranges from insulating to metallic $VO_2$ phase and a clear transition area occurs before the $VO_2$ critical temperature $T_C$. (b) Angular transmittance as a function of temperature. Both results are plotted at a fixed wavelength of 7.6μm.

## 3. Optical transmission switching

Transient simulations based on pump-probe laser excitations are performed to demonstrate the ability of the directional filter to operate as a fast optical transmission switch. The heating process in the simulations is induced by the pump laser excitations, where two different lasers operating at distinct wavelengths and input intensities are used. A detailed explanation of the transient simulations is available in the Supplementary Information [28]. Comsol Multiphysics® is used to simulate the thermal and optical behavior of the filter. The pump and probe lasers are modelled as plane waves which are launched towards the multilayer structure. Although typical laser pulses have a gaussian beam spatial profile, their spatial extend usually covers a large area. Hence, we focus our simulations near the maximum of the spatial gaussian beam distribution and approximate the impinging lasers as plane waves. The temperature of the structure is initialized to room temperature (293K) and the Comsol's electromagnetic heating interface is used to determine the heating of the structure according to the following equations:

$$Q_e = Q_{rh} + Q_{ml}, \quad (6)$$

$$Q_{rh} = \frac{1}{2}Re(\boldsymbol{J} \cdot \boldsymbol{E}^*), \quad (7)$$

$$Q_{ml} = \frac{1}{2}Re(i\omega\boldsymbol{B} \cdot \boldsymbol{H}^*), \quad (8)$$

where $Q_e$ is the total heat added to the system from electromagnetic heating, $Q_{rh}$ is the heat from resistive losses, $Q_{ml}$ is the heat from magnetic losses, J is the current density, E is the induced electric field, ω is the angular frequency, and B and H are the induced magnetic flux density and field, respectively. For each time step, the temperature of the filter is first determined using Eqs. (6-8) and then the optical constants of $VO_2$ are calculated using Eq. (4). The optical constants of $VO_2$ are then plugged into an electromagnetic simulation to determine the reflectance and transmittance of the filter. The probe laser is launched as a monochromatic plane wave and the power flow through the structure is measured. The calculated transmittance is then recorded for each time step. A simple schematic of the pump-probe simulation setup is shown in the inset of Fig. 4d.

    The first choice of pump laser is a Thulium (Tm) doped fiber laser operating at a wavelength of 1940nm with intensities of 5, 7.5, and 10mW/cm$^2$ [31]. The second pump choice is an yttrium vanadate (Nd:YVO$_4$) laser operating at 532nm with slightly lower intensities of 2.5, 5, and 7.5mW/cm$^2$ [32]. The chosen pump lasers are commercially available continuous wave (CW) lasers, but their excitation is chosen to last for 5ms to allow the structure to cool

and demonstrate its transmission switching behavior during the different phases. The filter is always illuminated by a normal incident low power laser source acting as probe with fixed wavelength (7.6μm) in the mid-IR range that does not cause additional heating. The computed induced temperature along the filter as a function of time for the two pump lasers is demonstrated in Figs. 4a and 4b, respectively. The filter is initially held at room temperature (293K). Heating is practically uniform throughout the structure, and the computed temperature represents the average temperature through the entire filter geometry. The induced temperature exceeds the critical temperature (purple dashed line in Figs. 4a and 4b) associated to the $VO_2$ phase transition, as was described in Fig. 3. This phase transition leads to a transient and relatively abrupt switching in the transmittance of the filter for each probe laser which is depicted in Figs. 4c and 4d.

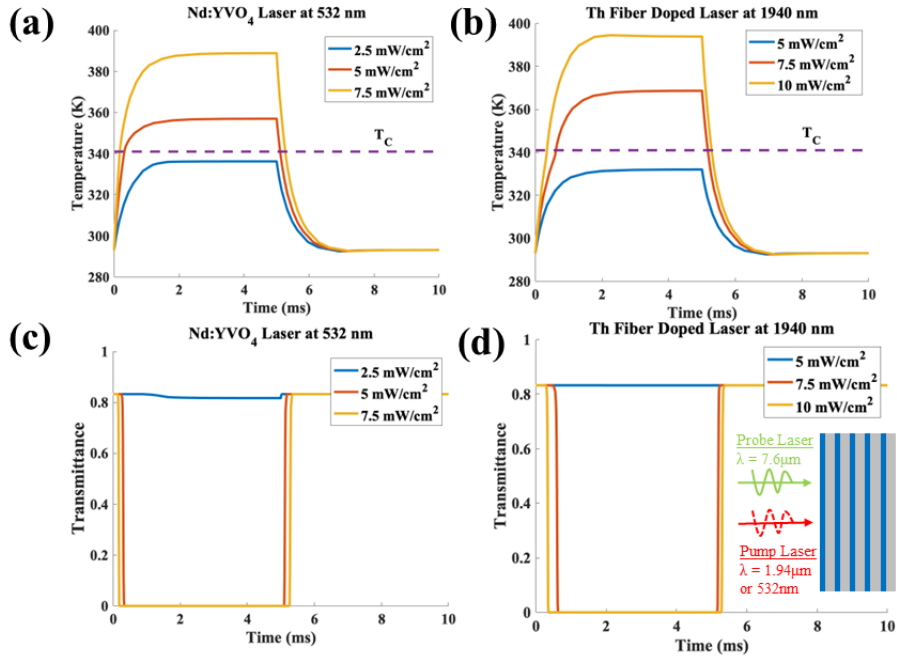

Fig 4. (a)-(b) Temperature and (c)-(d) transmittance of the directional filter as a function of time for the (a), (c) Nd:YVO$_4$ and (b), (d) Th doped fiber pump laser, respectively. The transient results are presented for different pump laser intensities. (Inset in (d)) A simple schematic of the simulation setup showing the pump/probe scheme.

The process of transmission switching takes less than a millisecond with both currently used pump lasers. More specifically, it is derived from Fig. 4 that the insulating to metallic phase change occurs at approximately 635μs and 361μs for the fiber laser at 7.5 and 10 mW/cm$^2$ intensities, respectively. The transition from the metallic back to the insulating state takes 240μs and 368μs for the same laser with similar intensities. In the case of the yttrium vanadate laser, the insulator-metal transition occurs at 335μs and 186μs while the metal-insulator transition happens at 176μs and 335μs for intensities of 5 and 7.5mW/cm$^2$ respectively. We define the insulator-metal transition time as the time it takes for the transmission to drop to zero after the start of the pump laser pulse. The metal-insulator transition time is defined as the time it takes for the transmission to increase back to maximum after the pump laser pulse is removed. The yttrium vanadate laser has faster switching performance mainly due to the stronger absorption of insulating $VO_2$ in the visible emitting wavelength of this pump laser. The filter's switching process can become even faster, approaching few micro- or pico-seconds, if less layers (see Supplementary Information [28]) or ultrafast pulsed lasers are used, respectively, resulting to a more rapid heating process [33-35]. The results shown here are very fast compared to other relevant schemes, such as liquid-based switching, which had a ~200ms response time [18], or liquid crystal designs that had switching times on the order of milliseconds [19].

## 4. Conclusions

We presented the design of a new thermally controlled directional narrowband optical transmission filter. The use of $VO_2$ enables tunable and fast switching response between high and zero transmission. The transmission through the filter is limited to narrow bands in the mid-IR range and only light near normal incidence is allowed to pass through the filter. The filter works for both polarizations and, as a result, can operate with incoherent light sources, such as thermal radiation. The proposed optical filter will have applications in infrared optical communications, where its switching capability and directional selective transmission can lead to a reconfigurable and tunable response [36].

**Funding.** This work was partially supported by the Office of Naval Research Young Investigator Program (ONR-YIP) (Grant No. N00014-19-1-2384), the National Science Foundation/EPSCoR RII Track-1: Emergent Quantum Materials and Technologies (EQUATE) (Grant No. OIA-2044049), and the NASA Nebraska Space Grant Fellowship.

# Supplementary Material

## Tunable Directional Filter for Mid-Infrared Optical Transmission Switching

Andrew Butler, Jack Schulz, and Christos Argyropoulos[*]

Department of Electrical and Computer Engineering, University of Nebraska-Lincoln, Lincoln, Nebraska 68588, USA

*christos.argyropoulos@unl.edu

**Transient simulation details**

The Finite Element Method (FEM) full wave simulation software Comsol Multiphysics® was used to analyze the transient behavior of the presented vanadium dioxide ($VO_2$) directional filter. Figure S1 shows a schematic of the electromagnetic simulation set-up. Port boundary conditions were used to launch either transverse magnetic (TM) or transverse electric (TE) polarized plane waves. The side boundaries were simulated using periodic boundary conditions. The $VO_2$ and calcium fluoride ($CaF_2$) materials were modelled using their frequency dielectric constants [1,2].

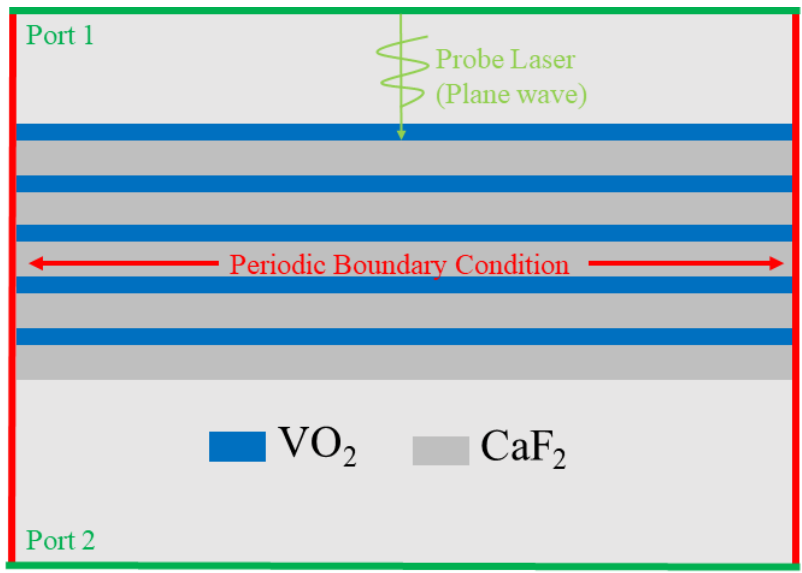

Fig S1. Schematic of the electromagnetic simulation and boundary conditions.

A pump-probe laser setup was used to model the transient heating process and calculate the transmittance as a function of time. Both pump and probe lasers were launched from the port 1 boundary. The probe laser wavelength and input intensity were selected to resemble realistic laser parameters. Two lasers were used as probes in the modeling, where the first is a Thulium (Tm) doped fiber continuous wave (CW) laser operating at 1940nm and the second is a yttrium vanadate (Nd:YVO$_4$) CW laser emitting at 532nm. The input power intensity was varied between 2.5-10mW/cm$^2$. Typical lasers have a Gaussian beam spatial profile, but the small area of the simulation geometry allowed us to approximate the laser as a plane wave. The probe CW lasers were chopped at 5ms duration to demonstrate the effect of VO$_2$ phase transition in the transmittance of the filter.

The temperature of the filter was initialized to room temperature (293K) and was set uniform throughout the structure at this initial point. The COMSOL's electromagnetic heating interface was used to introduce a heat source term to the heat interface according to the following equations:

$$Q_e = Q_{rh} + Q_{ml}, \qquad (1)$$

$$Q_{rh} = \frac{1}{2} Re(\mathbf{J} \cdot \mathbf{E}^*), \qquad (2)$$

$$Q_{ml} = \frac{1}{2} Re(i\omega \mathbf{B} \cdot \mathbf{H}^*), \qquad (3)$$

where $Q_e$ is the total heat added to the system from electromagnetic heating, $Q_{rh}$ is the heat from resistive losses, $Q_{ml}$ is the heat from magnetic losses, $\mathbf{J}$ is the current density, $\mathbf{E}$ is the induced electric field, $\omega$ is the angular frequency, and $\mathbf{B}$ and $\mathbf{H}$ are the induced magnetic flux density and field, respectively. Since the materials at the currently studied frequency range are nonmagnetic, Eq. (3) can be neglected, and heating occurs entirely due to the electric field. A schematic of the heating simulation is shown in Fig. S2. Periodic conditions are imposed on the side boundaries of the model, and an ambient temperature of 293K (room temperature) is applied to the top and bottom boundaries. VO$_2$ and CaF$_2$ are modelled in the thermal simulations using their thermal conductivity, heat capacity, and density values taken from experimental data [4,5].

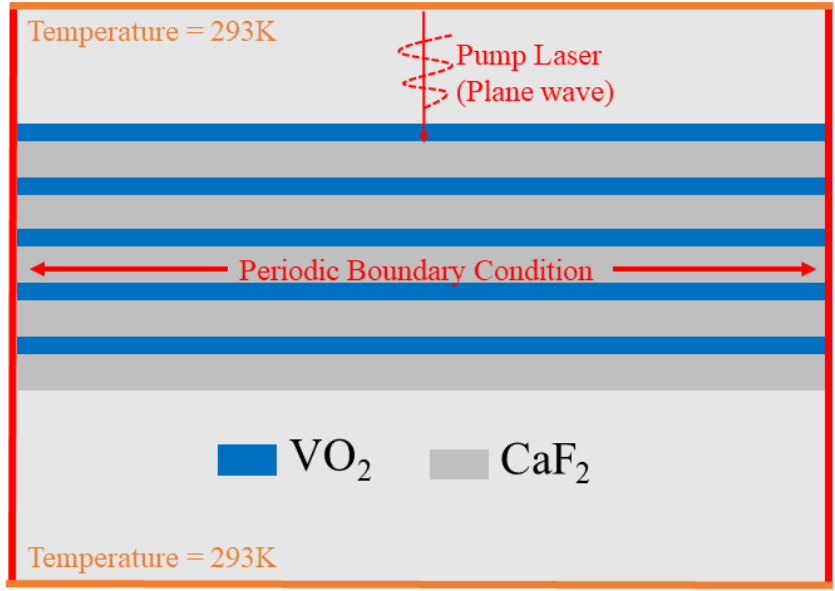

Fig. S2 Schematic of the thermal simulation and boundary conditions.

Once the temperature of the structure was determined, the $VO_2$ dielectric constant was calculated using Eqs. (4)-(5) in the main paper and the pump laser was launched from the port boundary. The pump laser was modelled as a TM polarized plane wave with 7.6µm wavelength but similar response is obtained for TE polarization, as depicted in Fig. 1d in the main paper. The S-parameter calculations were performed to calculate the reflection and transmission coefficients of the structure and the port boundaries measured the total reflected and transmitted power flow through the structure. The power flow through each port was calculated using:

$$P = \int_C \vec{S} \cdot \vec{n} = \frac{1}{2}\int_C Re\{\vec{E} \times \vec{H}^*\} \cdot \vec{n}, \qquad (4)$$

where $\vec{S}$ is the time averaged Poynting Vector, C is the curve of the port boundary, $\vec{n}$ is the normal vector, $\vec{E}$ is the electric field vector, and $\vec{H}$ is the magnetic field vector. The S-parameters were then calculated as:

$$S_{11} = \sqrt{\frac{power\ reflected\ back\ to\ port\ 1}{power\ emitted\ from\ port\ 1}}, \qquad (5)$$

$$S_{21} = \sqrt{\frac{power\ transmitted\ to\ port\ 2}{power\ emitted\ from\ port\ 1}}. \quad (6)$$

The reflectance and transmittance were finally computed as $R = |S_{11}|^2$ and $T = |S_{21}|^2$, respectively.

**Effect of number of layers**

The effect of reducing the number of unit cells or layers in the filter is demonstrated in Fig. S3. Figures S3a, S3b, and S3c demonstrate the transmittance as a function of angle for 3, 4, and 5 unit cells, where each unit cell is composed of one layer of $VO_2$ and one layer of $CaF_2$. The spectral transmittance comparison of the same filter for 3, 4, and 5 unit cells is demonstrated in Fig S3d for the normal incidence case. Finally, the angular transmittance of the filter for 3, 4, and 5 unit cells is shown in Fig. S3e. Decreasing the number of unit cells widens the transmission bandwidth of the filter, reduces directionality, and slightly increases transmittance. The effect on the filter performance is not drastic, however for some applications reducing the number of unit cells could help reduce cost and simplify fabrication.

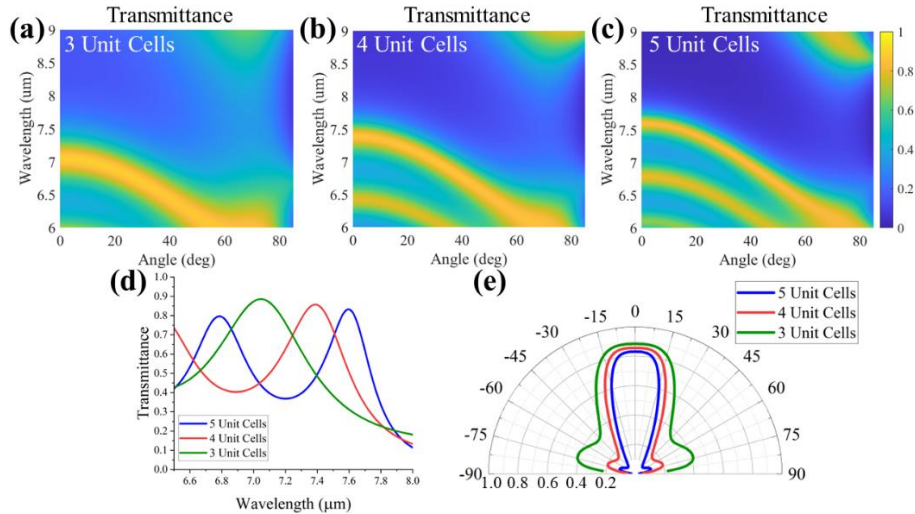

Fig. S3: Transmittance as a function of wavelength and incident angle for (a) 3, (b) 4, and (c) 5 unit cells, respectively. (d) Transmittance spectrum for 3, 4, and 5 unit cells at normal incidence illumination. (e) Directional transmittance pattern as a function of angle for 3, 4, and 5 unit cells. The wavelengths used in this plot are selected to coincide with each design's maximum transmission point.

Figure S4 shows the transient simulation results for 3 unit cells compared to the 5 unit cell case which is presented in Fig. 4 in the main paper. More specifically, Figs. S4a and S4b depict the temperature as a function of time while Figs. S4c and S4d demonstrate the transmittance as a function of time for the simulated yttrium vanadate (Nd:YVO4) and Thulium (Tm) doped fiber probe lasers. Pump laser intensities of 7.5mW/cm$^2$ and 10mW/cm$^2$ are used for these simulations for both yttrium vanadate and Tm doped fiber lasers, respectively. Reducing the number of unit cells decreases the phase switching times in the case of the yttrium vanadate laser. The insulator-metal transition time is reduced from 186μs to 139μs and the metal-insulator transition time is reduced from 335μs to 56μs. This is because there is less material to be heated and the heat diffuses throughout the entire filter more quickly. However, for the case of the Tm doped fiber laser, the insulator-metal transition time is increased from 361μs to 661μs and the metal-insulator transition time is reduced from 368μs to 152μs. This is due to the weaker absorption of insulating $VO_2$ at the Tm doped laser's wavelength. Reducing the number of $VO_2$ layers decreases the overall filter absorption which, subsequently, slows the heating process. On the other hand, reducing the size of the filter enables faster cooling, which decreases the metal-insulator transition time. Once the $VO_2$ becomes metallic, the temperature increases quickly because the absorption is always higher in the metallic phase. Finally, the choice of probe laser is important for the heating process due to the distinct wavelength emitted from each probe laser that will interact in a different way with the geometrical features of each layer along the filter structure.

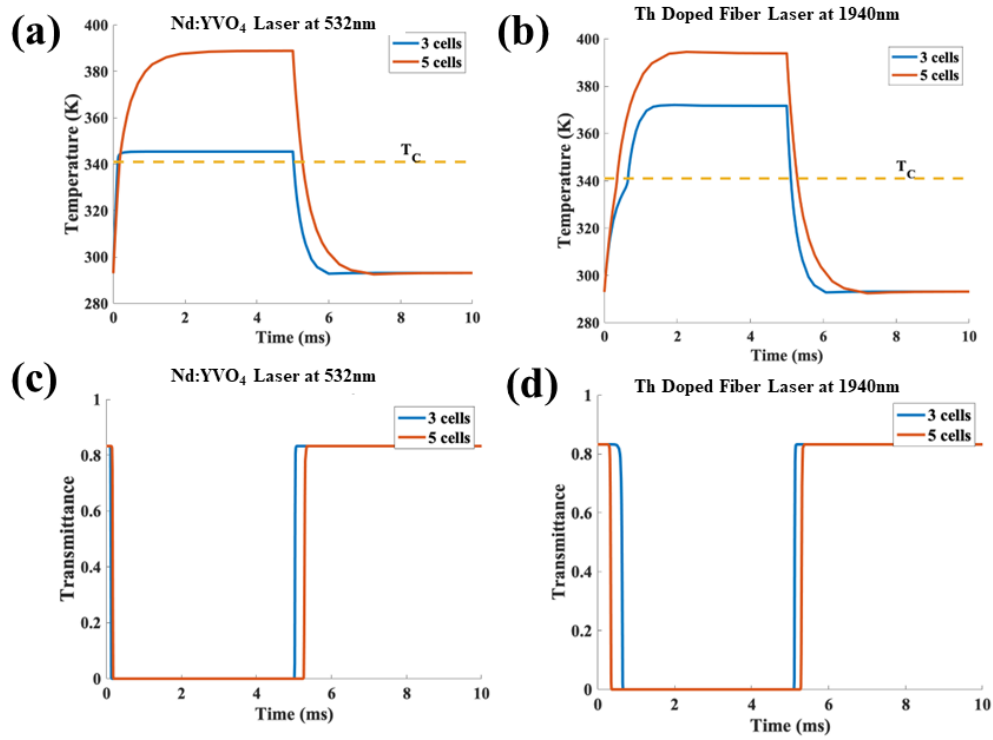

Fig. S4: (a)-(b) Temperature and (c)-(d) transmittance of the directional filter with 3 and 5 unit cells as a function of time for the (a), (c) Nd:YVO$_4$ and (b), (d) Th doped fiber pump laser when illuminated by fixed intensities of 7.5mW/cm$^2$ and 10mW/cm$^2$, respectively.

**Reflection and absorption in the metallic state**

In the metallic state, the transmittance of the filter drops to zero and the reflectance and absorptance are increased. Figure S5 shows the absorptance and reflectance of the filter as a function of angle and wavelength for VO$_2$ in the metallic state. At angles near normal incidence, the filter exhibits mirror-like response with some absorption, while for higher angles of incidence the device becomes a broadband absorber. This defines two extra functionalities of the proposed filter design as a reflector and a directional broadband absorber.

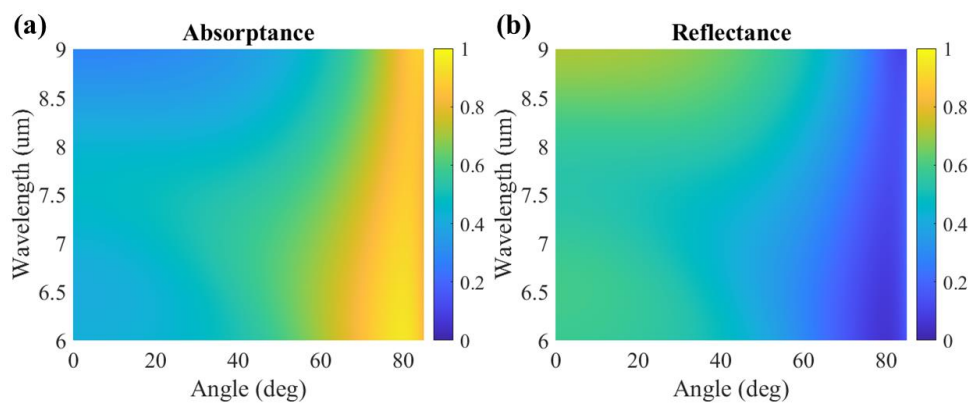

Fig. S5. (a) Absorptance and (b) reflectance as a function of incident angle and wavelength for the filter when $VO_2$ is in the metallic state.